\documentclass[twocolumn]{aastex62}

\graphicspath{{./}{figures/}}
\usepackage{gensymb}
\usepackage{amsmath}
\usepackage{appendix}
\usepackage{flafter}

\newcommand{\sub}{MAXI~J1820}

\newcommand{\hx}{\it Insight-HXMT}
\newcommand{\hbAppendixPrefix}{A}

\shorttitle{Broadband temporal features in MAXI~J1820+070}
\shortauthors{Wang et al.}

\begin{document}

\title{
\Large \textbf{
The evolution of the broadband temporal features observed in the black-hole transient MAXI~J1820+070 with {\hx}}
}

\author{Yanan Wang} 
\affil{Universit\'e de Strasbourg, CNRS, Observatoire astronomique de Strasbourg, UMR 7550, F-67000 Strasbourg, France}
\altaffiliation{E-mail: yanan.wang@astro.unistra.fr}

\author{Long Ji} 
\affil{Institut für Astronomie und Astrophysik, Kepler Center for Astro and Particle Physics, Eberhard Karls Universität, Sand 1, D-72076 Tübingen, Germany}

\author{S.~N. Zhang}
\affil{Key Laboratory for Particle Astrophysics, Institute of High Energy Physics, Chinese Academy of Sciences, 19B Yuquan Road, Beijing 100049, China}
\affil{University of Chinese Academy of Sciences, Chinese Academy of Sciences, Beijing 100049, China}

\author{Mariano M\'endez} 
\affil{Kapteyn Astronomical Institute, University of Groningen, PO Box 800, 9700 AV Groningen, The Netherlands}

\author{J.~L. Qu}
\affil{Key Laboratory for Particle Astrophysics, Institute of High Energy Physics, Chinese Academy of Sciences, 19B Yuquan Road, Beijing 100049, China}
\affil{University of Chinese Academy of Sciences, Chinese Academy of Sciences, Beijing 100049, China}

\author{Pierre Maggi} 
\affil{Universit\'e de Strasbourg, CNRS, Observatoire astronomique de Strasbourg, UMR 7550, F-67000 Strasbourg, France}

\author{M. Y. Ge}
\affil{Key Laboratory for Particle Astrophysics, Institute of High Energy Physics, Chinese Academy of Sciences, 19B Yuquan Road, Beijing 100049, China}

\author{Erlin Qiao} 
\affil{Key Laboratory of Space Astronomy and Technology, National Astronomical Observatories, Chinese Academy of Sciences, Beijing 100101, China}

\author{L. Tao}
\affil{Key Laboratory for Particle Astrophysics, Institute of High Energy Physics, Chinese Academy of Sciences, 19B Yuquan Road, Beijing 100049, China}

\author{S. Zhang}
\affil{Key Laboratory for Particle Astrophysics, Institute of High Energy Physics, Chinese Academy of Sciences, 19B Yuquan Road, Beijing 100049, China}

% ...

\date{Accepted ? December ?. Received ? December ?; in original form ? December ?}

\begin{abstract}
We study the evolution of the temporal properties of MAXI~J1820+070 during the 2018 outburst in its hard state from MJD~58190 to 58289 with {\hx} in a broad energy band 1--150~keV.
We find different behaviors of the hardness ratio, the fractional rms and time lag before and after MJD~58257, suggesting a transition occurred at around this point. 
The observed time lags between the soft photons in the 1--5~keV band and the hard photons in higher energy bands, up to 150~keV, are frequency-dependent: the time lags in the low-frequency range, 2--10~mHz, are both soft and hard lags with a timescale of dozens of seconds but without a clear trend along the outburst; the time lags in the high-frequency range, 1--10~Hz, are only hard lags with a timescale of tens of milliseconds; first increase until around MJD~58257 and decrease after this date. The high-frequency time lags are significantly correlated to the photon index derived from the fit to the quasi-simultaneous {\it NICER} spectrum in the 1--10~keV band. This result is qualitatively consistent with a model in which the high-frequency time lags are produced by Comptonization in a jet.

\end{abstract}


\section{Introduction} \label{intro}
X-ray variability is present in accreting black-holes (BHs) on timescales of milliseconds to years: for example, flares \citep{Belloni2000,Altamirano2011}, quasi-periodic oscillations (QPOs, \citealt{Klis1989}), dips \citep{Kuulkers1998,Kajava2019} and broadband noise \citep{Takizawa1997,Mendez1997b,Casella2005,Motta2015}.
The variability associated with source spectral state reveals not only the mass accretion rate but also the geometry and structure of the system. 
Understanding the variability is crucial both to understanding the accretion process and to determining the source properties \citep{Uttley2014}.

Another important measurement of the variability is the time lag between soft and hard photons (e.g. \citealt{Nowak1999,Altamirano2015,Zhang2017}). The time-scale of time lags reflects the underlying physical process that produces the variability and the evolution of the lags along the outburst may also reveal the changes of the geometry of the system. Hard lags (hard photons lag the soft ones) correlated with the shape of the spectral continuum up to around 30~keV in the hard state have been reported in several systems, e.g. Cyg~X--1 \citep{Pottschmidt2003,Grinberg2014} and GX~339--4 \citep{Nowak2002,Altamirano2015}. \cite{Reig2018} analyzed this correlated behavior in twelve outbursts of eight BH systems and confirm the photon-index–time-lag correlation as a global property of BH X-ray binaries.

Additionally, three main canonical states driven by the mass accretion rate have been widely used to describe a full outburst in black-hole transients: hard, soft and intermediate states, identified by their timing and spectral properties \citep{Tanaka1995,Klis1995,Mendez1997a,Remillard2006}. 
Each state shows timing and spectral complex characteristics; hardness, defined as the ratio between count rates in different bands, has been demonstrated as a useful parameter to understand the spectral evolution of outbursts. 
However, one sometimes failed to finish a transition loop; for the transients that do not enter a soft state, this type of outburst has been dubbed `failed' outburst, e.g. H~1743--322 \citep{Capitanio2009,Zhou2013} and Swift~J1753.5--0127 \citep{Soleri2013}. \cite{Capitanio2009} proposed that the case during the 2008 outburst of H~1743--322 was associated with a premature decrease of the mass accretion rate.

The black-hole candidate (BHC) MAXI~J1820+070 was discovered on 2018 March 11 in X-rays with the {\it Monitor of All-sky X-ray Image} ({\it MAXI}, \citealt{Matsuoka2009}). After four days, the {\it Hard X-ray Modulation Telescope} ({\it HXMT}), dubbed {\hx}, started obtaining observations, which extended until October 2018. {\sub} is a bright source with a luminosity of up to $10^{37}~\rm erg~s^{-1}$ in the energy band 0.01--100~keV \citep{Shidatsu2019}, assuming a distance of 3~kpc \citep{Gandhi2018}. This nearby luminous X-ray source, brighter than 4~Crab in 15--50~keV, makes itself an ideal target for {\hx}. Recently, \cite{Torres2019} confirmed {\sub} to be a black-hole transient with a dynamical mass measurement, constraining the BH mass to be 7--8~$M_{\rm \odot}$ when the inclination angle is $69-77 \degree$. Later on, \cite{Atri2019} refined the mass of the black hole in {\sub} to be $9.5\pm1.4~M_{\rm \odot}$ by measuring the distance to the source with the Very Long Baseline Array and the European Very Long Baseline Interferometry Network.

\cite{Kara2019} conducted spectral and timing analysis to study the geometry of the hard emission region in {\sub}.
By observing the changes of the reverberation lags between 0.1--1.0~keV and 1.0--10.0~keV energy photons, they suggest that the corona shrinks vertically as the source evolves from the hard towards the soft state. Meanwhile, the accretion disk shows no evolution along the outburst. Both findings have been supported by the work done by \cite{Buisson2019} in which they analyzed some {\it Nuclear Spectroscopic Telescope Array} ({\it NuSTAR}, \citealt{Harrison2013}) observations, covering the period of the data used in \cite{Kara2019}.

In this paper, we carry out a detailed time- and Fourier- domain analysis of {\sub} using {\hx} data, together with the spectral analysis using {\it NICER} data, to investigate the X-ray variability observed in the luminous hard state.

\section{Observations and data reduction}
{\hx} was launched on 2017 June 15. It carries three slat-collimated instruments on board \citep{Zhang2020}, the High Energy X-ray telescope (HE: 20--250~keV, \citealt{Liu2020}), the Medium Energy X-ray telescope (ME: 5--30~keV, \citealt{Cao2020,Guo2020}) and the Low Energy X-ray telescope (LE: 1--15~keV, \citealt{Chen2020,Liao2020}).

The entire outburst of {\sub} has been observed with {\hx} between March and October 2018.
Owing to the broad energy band and high time resolution (LE: 1~ms; ME: 280~$\mu$s; HE: 25~$\mu$s) of {\hx}, we performed a statistical analysis of the X-ray timing properties of {\sub}, up to 150~keV, of 63 observations from MJD~58197 to 58288, in the hard state of the outburst (see the definition of the spectral states in \citealt{Shidatsu2019}). 
In Fig.~\ref{fig:lc}, we show the 15--50~keV long-term {\sc Swift}/BAT light curve with the indication of the spectral states of {\sub}. The red triangles and circles in panel~a correspond to the simultaneous observational time of {\hx}. 

We extracted the data from all three instruments using the {\hx} Data Analysis software ({\sc hxmtdas}) v2.00 \footnote{http://www.hxmt.org/index.php/dataan/fxrj}. 
We created the good-time-intervals based on the suggested criteria: (1) the offset for the pointing position is $\leq0.05 \degree$; (2) the Earth elevation angle is $>6 \degree$; (3) the geomagnetic cutoff rigidity is $>6 \degree$; (4) the extraction time is at least 300~s before or after the South Atlantic Anomaly (SAA) passage. To avoid possible contamination from the bright Earth and nearby sources, we applied the small field of view (FoV) for all the detectors \citep{Chen2018,Huang2018}.

We conducted spectral analysis with {\it NICER} in the 1--10~keV of {\sub}. We chose {\it NICER} over {\hx} because {\it NICER} has a larger effective area than {\hx} in this energy band.
The data processing and filtering was performed using HEASoft 6.26.1 and NICERDAS version 6. 
We followed standard data reduction steps to run the filtering, calibration and merging of {\it NICER} events. We excluded two of the active FPMs~14 and 34 which are often found to exhibit episodes of increased detector noise.
After data had been cleaned and calibrated, we extracted the {\it NICER} spectra using {\sc xselect}. The background is estimated with the tool, {\sc nibackgen3C50}.

\section{Results}
\begin{figure*}
    \centering
    \includegraphics[width=0.65\textwidth]{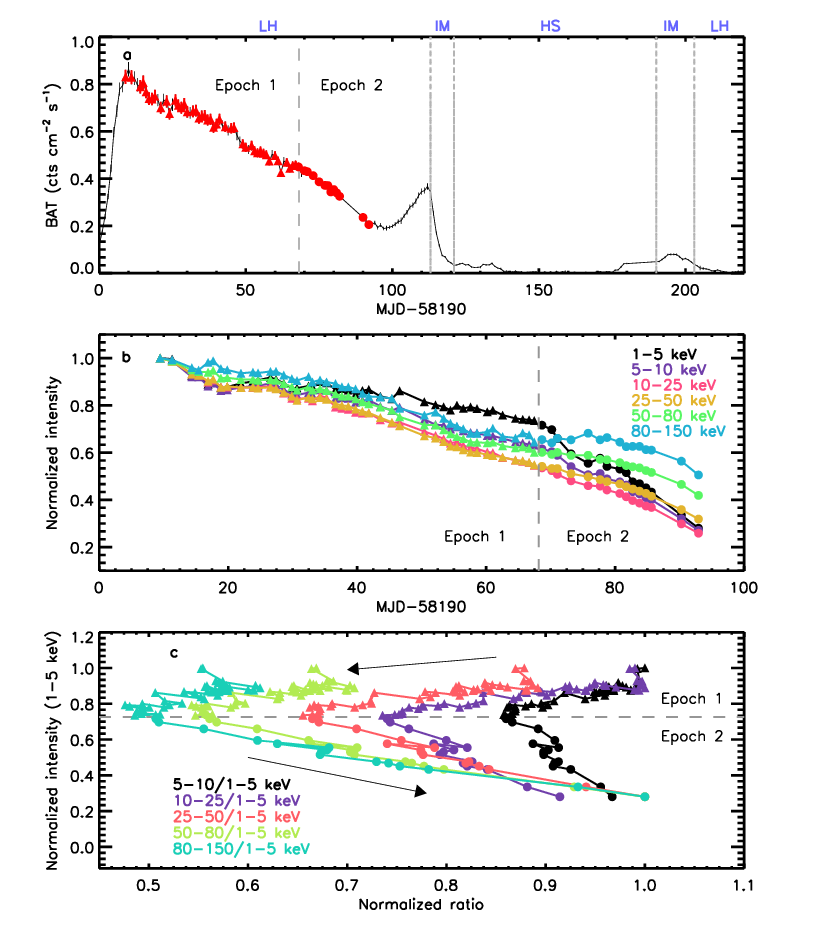} 
        \caption{Panel~a: The {\it Swift}/BAT lightcurve of {\sub} in the 15--50~keV band during the 2018 outburst. The red triangles and circles, corresponding to epoch~1 and 2 (here and in the following figures), respectively, indicate the simultaneous {\hx} observations used in this paper. The gray dotted lines indicate the states of the source, taken from \protect{\cite{Shidatsu2019}}. `LH', `IM', `HS' are short for low hard, intermediate, and high soft state, respectively. 
        Panel~b: The normalized {\it HXMT} lightcurves in different energy bands. 
        Panel~c: the normalized HID in different energy bands, evolving from the top right to the left and back to the bottom right as indicated by the arrows. The black, purple, orange, green and cyan symbols indicate the hardness ratio between, respectively, the energy bands 5--10~keV, 10--25~keV, 25--50~keV, 50--80~keV and 80--150~keV, with respect to the 1--5~keV band.}  
\label{fig:lc}
\end{figure*}

The evolution of the BAT light curve in Fig.~\ref{fig:lc}a shows that the intensity of {\sub} increased rapidly by a factor of $\sim7$ and reached the peak of the outburst at $\rm \sim0.85~BAT~counts~s^{-1}~cm^{-2}$ within the first 11~days, from MJD~58190 to 58200. In the following 90~days the intensity slowly decreased to around $\rm 0.2~BAT~counts~s^{-1}~cm^{-2}$. However, the intensity of {\sub} increased again on MJD~58288 and then reached a peak two times fainter than the previous one. At around MJD~58306 the outburst reached a local minimum of the intensity. 

We generated the {\hx} light curves in the 1--5~keV, 5--10~keV, 10--25~keV, 25--50~keV, 50--80~keV and 80--150~keV energy bands to see how the X-ray emission in different energy bands changes in the luminous hard state. We then computed hardness ratios using the 1--5~keV light curve as the reference. We show the normalized light curves and hardness-intensity diagrams in Fig.~\ref{fig:lc} in which we divided the intensity and the hardness ratio by their maximum values in each energy band for displaying purpose.
Fig.~\ref{fig:lc}b shows that the intensity in all energy bands decreases with time; the intensity in the softest band, 1--5~keV, shows the slowest decrease before MJD~58257 and the fastest decrease after MJD~58257. 
Fig.~\ref{fig:lc}c shows that the hardness ratio first decreases along the outburst, reaching a local minimum at around MJD~58246, then remains constant for a while and finally increases from MJD~58257. The evolution of the HIDs suggests a transition at MJD~58257. 
We therefore separated the dataset into two groups, epoch~1 and 2, before and after the transition point. The triangles and circles represent respectively epoch~1 and 2 in all figures in this work.

\subsection{X-ray variability in time- and Fourier-domain}
\begin{figure}
    \centering
    \vspace{2px}
    \mbox{\includegraphics[width=1\linewidth]{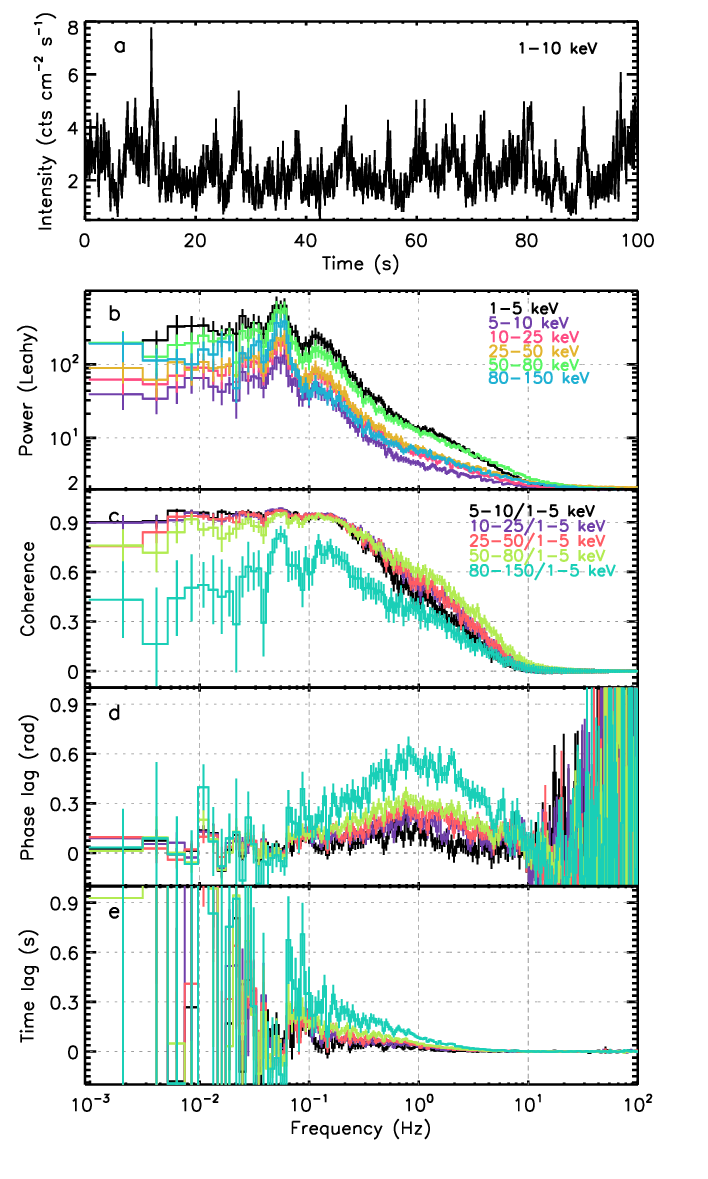}}
        \caption{From top to bottom: the power spectrum, raw coherence, phase lags and time lags in different energy bands of {\sub} at MJD~58200.} 
\label{fig:lc_pds}
\end{figure}

Fig.~\ref{fig:lc_pds}a shows part of the 1--10~keV light curve of the black hole transient {\sub} at MJD~58200. Similar to some other accreting BHs (e.g. GX~339--4 and XTE~J1550--564), the light curve of {\sub} shows various flares, which appear as peaked broad band noise in Fourier power density spectrum (PDS). To explore this further, we created power spectra with stingray\footnote{https://github.com/StingraySoftware/stingray}, a software for timing analysis of X-ray data \citep{Huppenkothen2019}: For this, we divided the data into continuous 500-s segments with a time bin of 0.005~s, computed periodograms for each segment and averaged all periodograms in each observation.

Fig.~\ref{fig:lc_pds}b shows the corresponding power spectrum at MJD~58200 with Leahy normalization in the frequency range 0.002--100~Hz in different energy bands, while the power being constant at around 2 at frequencies above 100~Hz. We thus subtracted the Poisson noise level of 2 to calculate the fractional root-mean-square (rms).
The power spectra in the 1--5~keV and 5--10~keV bands, respectively, show the highest and lowest power; the power spectra in other energy bands rank in between those two.
Each of power spectrum can be described by a combination of several Lorentzians and a power law, displaying broadband noise plus a low-frequency QPO with a centroid frequency at 0.054~Hz. The QPO centroid frequency in different energy bands shows no significant shift.
A second peak in the PDS is potentially a harmonic of the 0.054~Hz QPO, although data with higher signal to noise would be required to confirm this.

\begin{figure}
    \centering
    \mbox{\includegraphics[width=1\linewidth]{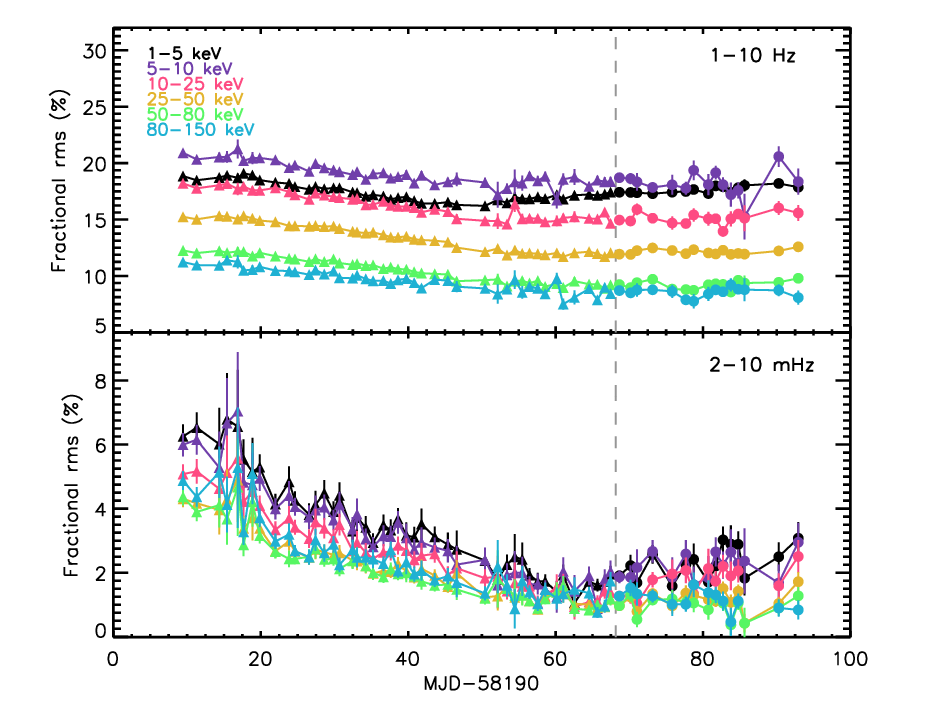}} \hspace{-8px}
        \caption{Frequency-dependent fractional rms amplitude of the broadband noise in {\sub} in different energy bands. } 
\label{fig:frac_rms}
\end{figure}

The centroid frequency of the detected QPOs in the observations used in this work ranges between 0.04 and 0.56~Hz; the QPO frequency first increases with time in epoch~1 and then decreases with time in epoch~2. More details of the QPO information will be presented in Ma et al. (in prep).
As we aim to study the properties of the broadband noise, we computed the fractional rms in the low frequency range, 2--10~mHz, below the QPO centroid frequency, and in the high frequency range, 1--10~Hz, above the QPO centroid frequency, one measurement per observation.
We show the fractional rms as a function of time in Fig.~\ref{fig:frac_rms}. 

As shown in Fig.~\ref{fig:frac_rms}, the high-frequency fractional rms amplitude is larger than the low-frequency one in the same energy band;
the former slightly decreases along the outburst until MJD~58240 in epoch~1 and after that remains more or less constant; the latter shows a similar behavior to the former but with larger uncertainties, it decreases much more dramatically with time in epoch~1, and then increases with volatility in epoch~2 except for the rms at energies above 25~keV. The difference between the same energy bands of the former is larger than that of the latter. These results are consistent with the fact that the low-frequency part of the PDS changes more significantly than the high-frequency part along the outburst.

The largest fractional rms amplitude in the 1--10~Hz frequency range is in the 5--10~keV band, followed by the rms in the other energy bands, suggesting a significant contribution of the reflection component to the variability. Excepting that band, the fractional rms amplitude in all other energy bands, both in the low-frequency and high-frequency ranges, decreases with increasing energy.

\subsection{Frequency-dependent time lag} 
We computed the time lag between soft and hard photons \citep{Nowak1999} to constrain models and emission geometry. We first used stingray to create cross-spectra of each selected observation of {\sub} between the soft X-rays in the 1--5~keV band and the hard X-rays in the 5--10~keV, 10--25~keV, 25--50~keV, 50--80~keV and 80--150~keV bands, respectively. 
We illustrate the raw coherence, phase-lag and time-lag spectrum in Figs.~\ref{fig:lc_pds}c, d and e. The mean count rate for each band is 751.1, 146.4, 316.2, 1440.0, 651.9 and 364.5~$\rm cts~s^{-1}$, respectively.

There are two local peaks in each coherence spectrum at the QPO fundamental and suspected harmonic frequencies in the power spectra; both peaks become more significant with increasing energy (see Fig.~\ref{fig:lc_pds}c).
Figs.~\ref{fig:lc_pds}d and e show that both the phase- and time-lag spectra below the QPO frequency are noisy; a dip appears at around the QPO frequency, 0.054~Hz, and the lags immediately increase after that. 
Meanwhile the time lags are correlated with energies, being larger at higher energies. 
We show a few examples of the time lag versus energy, together with their power spectra, in Fig.~\ref{fig:lag_obs} in which we use the average photon energy for each energy band.

\begin{figure}
    \centering
    \mbox{\includegraphics[width=1\linewidth]{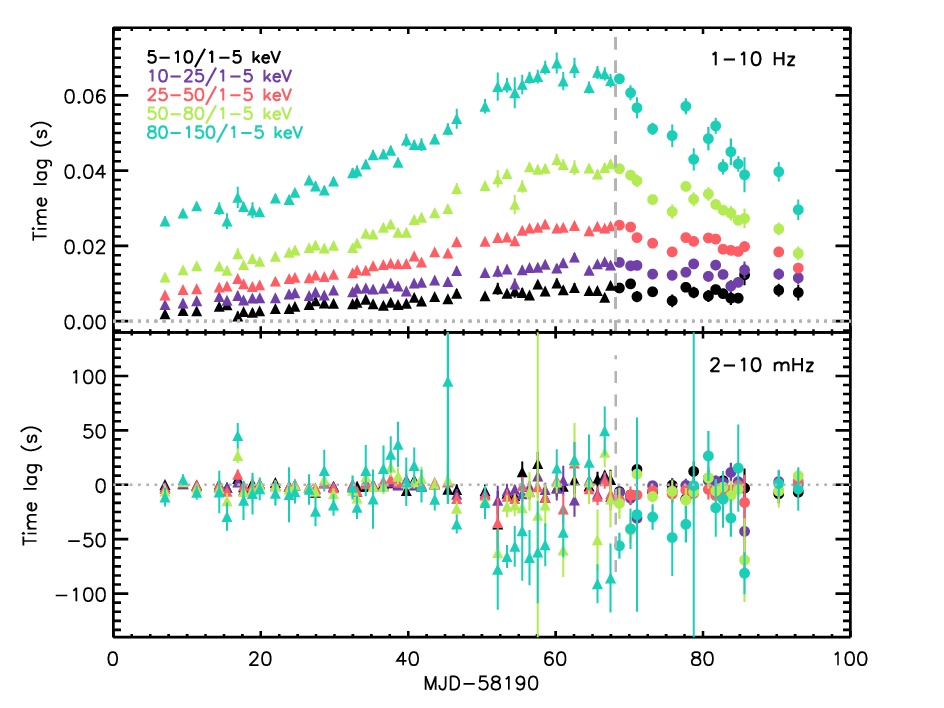}} 
        \caption{Frequency-dependent time lag of the broadband noise in {\sub} between the 1--5~keV and the 5--10~keV, 10--25~keV, 25--50~keV, 50--80~keV, 80--150~keV bands as a function of time.
        }  
\label{fig:lag}
\end{figure}

To study the evolution of the frequency-dependent time lags along the outburst, we calculated the time lags in the 2--10~mHz and 1--10~Hz frequency ranges per observation.
We first averaged the cross-spectrum over the 2--10~mHz and 1--10~Hz frequency ranges, respectively, and then calculated the time lags from the resulting phase lags of the averaged cross-spectrum.
As shown in the upper panel of Fig.~\ref{fig:lag}, the high-frequency time lags increase with energy; the time lags increase with time in epoch~1 and decrease with time in epoch~2, except that the one in the 1--5~keV band is more or less constant; a dip is present at MJD~58265 in all curves.
The low-frequency time lags, shown in the lower panel of Fig.~\ref{fig:lag}, are significantly larger than the high-frequency lags and show both positive and negative values but without a clear trend with time.

\subsection{Evolution of the spectral parameters  \label{sec:spec}}
Fig.~\ref{fig:lc} shows that the soft and hard parts of the spectrum evolve differently in epochs~1 and 2: the hard part of the spectrum drops more steeply than the soft part in epoch~1 whereas the soft part drops more steeply than the hard part in epoch~2, opposite to epoch~1; those changes result in decreased hardness ratio in epoch~1 and increased hardness ratio in epoch~2.
To trace the energy-dependent changes of the intensity, we studied the {\it NICER} spectra of {\sub} in the period of MJD~58198--58286 with exposure time longer than 1~ks. 
We analyzed the {\it NICER} spectra with the response matrix version 1.01 and the effective area file version 1.02\footnote{https://heasarc.gsfc.nasa.gov/docs/nicer/proposals/nicer\_tools.html}. 

We fit each spectrum in the band 1--10~keV with a three component model consisting of a hard component for the persistent emission ({\sc powerlaw} in XSPEC), a multi-temperature disk blackbody for the soft thermal emission ({\sc diskbb} in XSPEC) and a line model for the broad Fe emission line ({\sc gaussian} in XSPEC). 
To account for the interstellar absorption, in all fits we used the component {\sc tbabs} with solar abundances from \cite{Wilms2000} and photoelectric absorption cross-sections from \cite{Verner1996}.
The overall model is {\sc tbabs$*$(diskbb+gaussian+powerlaw)}.

Still, we found significant residuals at 2--3~keV which are likely instrumental due to Au. We applied Crab correction to the spectra as described in \cite{Ludlam2018}, whereas the residuals are still visible. We therefore ignored the energy band from 1.8 to 3~keV and added 0.5\% systemic error to the fit to minimize the instrumental effect. The spectra, before MJD~58290, require a narrow Fe emission line centered at 6.4--6.5~keV. Since either including this narrow line or not shows no significant effect on the best-fitting parameters of the continuum, we do not include this component in the fit. We quote errors at 90\% confidence level for the spectral analysis.

\begin{figure}
    \centering
    \mbox{\includegraphics[width=1\linewidth]{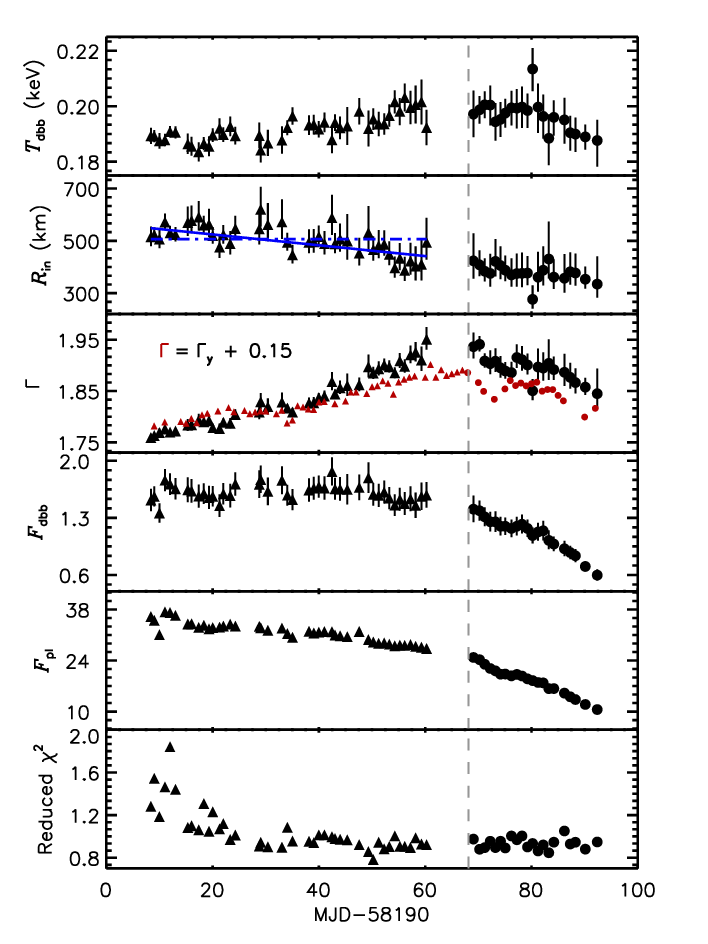}} 
        \caption{Best-fitting parameters of the {\it NICER} spectra of {\sub}. From top to bottom, the panels show, respectively, the evolution of the blackbody temperature, the inner radius of the accretion disk, the photon index, the {\sc diskbb} flux, the {\sc powerlaw} flux and the corresponding reduced $\chi^2$. $F_{\sc dbb}$ and $F_{\sc pl}$ are in units of $10^{-9}~\rm erg~s^{-1}~cm^{-2}$. The blue dash-dotted and solid lines separately represent the constant and linear fit to the $R_{\rm in}$-MJD relation in epoch~1. The photon index values plotted in red are from the paper of You et al. (in prep, see more details in Section.~\ref{sec:spec}).}  
\label{fig:spec}
\end{figure}

We show the temperature and the inner radius of the accretion disk, the power-law photon index and the reduced $\chi^2$ as a function of time in Fig.~\ref{fig:spec}. The inner radius is derived from the {\sc diskbb} normalization which is defined as $(R_{\rm in}/D_{\rm 10})^2*\rm cos \theta$, where $R_{\rm in}$ is the inner disk radius in km, $D_{\rm 10} $ is the distance to the source in units of 10~kpc and $\theta$ is the disk inclination angle. We assumed $D_{\rm 10}= 0.3$ \citep{Gandhi2018} and $\theta = 70\degree$ \citep{Torres2019} in this work. Both the disk temperature and the photon index increase, whereas the inner disk radius decreases in epoch~1, indicating that the source softens in this period.
We fitted the evolution of the inner disk radius with both a constant and a decreasing linear function and got a better fit with the latter one, with associated F-test probability $\sim 9 \times 10^{-8}$, indicating that the evolution of inner disk radius in epoch~1 decreased with time.

\begin{figure}
    \centering
    \mbox{\includegraphics[width=1\linewidth]{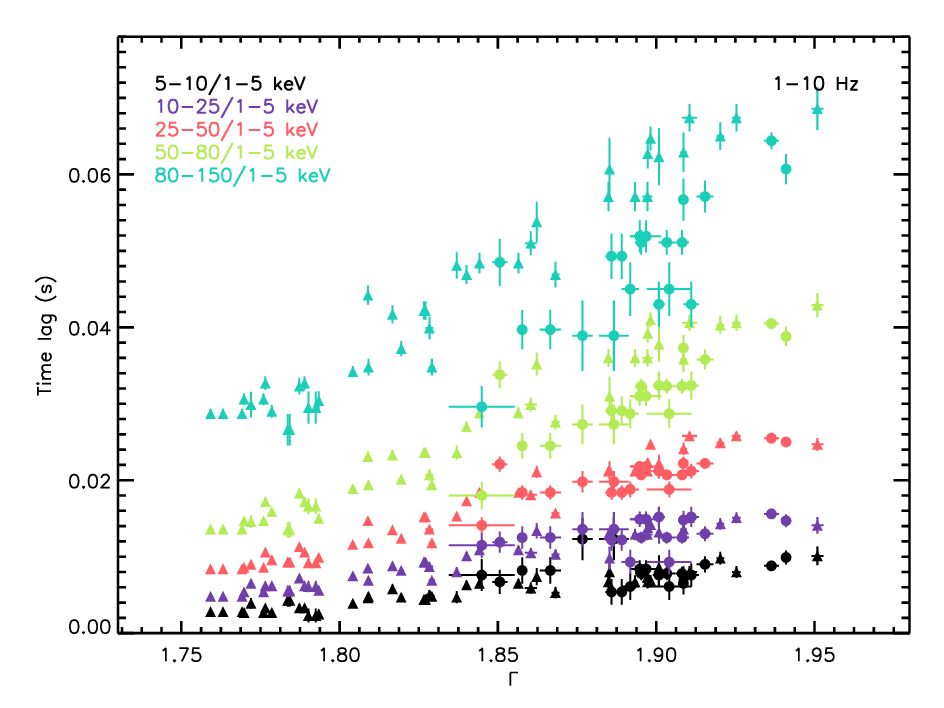}} 
        \caption{High-frequency time lags as a function of the photon index of the hard spectral component in {\sub}. As in Fig.~\ref{fig:lc}, the triangles and circles correspond to epoch~1 and 2, respectively.}  
\label{fig:lag_gamma}
\end{figure}

However, the source did not enter the soft state, and instead became harder in epoch~2: the disk temperature and the photon index decreased even though the inner disk radius remained constant.
The flux of the {\sc diskbb} component remains the same in epoch~1 when the flux of the {\sc powerlaw} component decreases by $\sim$30\%; both of the fluxes of the {\sc diskbb} and the {\sc powerlaw} components drop more than 50\% in epoch~2.
We plot the high-frequency time lags versus the photon index in Fig.~\ref{fig:lag_gamma} and find that these two quantities are significantly correlated in all selected energy bands. The slopes of lag-$\Gamma$ relation in epochs~1 and 2 are well consistent with each other except for the ones in the highest energy.

To test how much the assumption we made for the fit affects our result, we compare our best-fitting values of the photon index to the values taken from You et al. (in prep), since photon index is the only common parameter both in their and our work. 
They fit 199 {\hx} spectra from 3 to 150~keV with the model {\sc tbabs$*$(relxilllpCp+gaussian)}, in which {\sc relxilllpCp} \citep{Garcia2013} describes the reflection component plus the hard emission and {\sc gaussian} describes the narrow Fe emission line on top of the broad line in {\sc relxilllpCp}. We add their result to Fig.~\ref{fig:spec} as red points, in which we have offset the values by 0.15 for display purpose. 
Their power-law index values are smaller than ours which is likely due to the differences of the energy bands and the models applied by them and us. However, the overall trends of the evolution of the photon index in their and our work are consistent.

\section{Discussion}
We have studied the black hole candidate {\sub} the hard state of the 2018 outburst with {\hx} and {\it NICER} observations. By analyzing the variability in time and Fourier-domain, we found clear evidences of a transition occurring in this phase: the evolution of the hardness ratio, the fractional rms amplitude and the high-frequency time lag changed from epoch~1 to epoch~2. 
Additionally, we found that the high-frequency time lags are correlated with the photon index of the hard component in the energy spectrum.

\subsection{Hard-to-hard spectral state transition}
During the first decay of the 2018 outburst of {\sub}, we have observed a hard-to-hard transition which occurred at around MJD~58257, when we see the changes in the behavior of the hardness ratio, the fractional rms amplitude and the high-frequency time lags. Since both the hard and soft intensity decrease in this period, it appeared as if the source never entered into a soft state, we suppose that the source went though a failed outburst.
By observing the changes of the light curve and the frequency of the low-frequency QPOs in {\sub} with the {\it Swift}/XRT and {\it NICER} data, \cite{Stiele2019} also suggested that this source underwent a failed outburst.

A failed outburst is defined as one that either never leaves the hard state, or proceeds to an intermediate state, before returning to the hard state and quiescence during the outburst.
A number of cases suggest that the lack of soft-state transitions is possibly associated with a premature decrease of the mass accretion rate, as during the 2008 outburst of H1743--322 \citep{Capitanio2009}. 
On the other hand, \cite{DelSanto2016} reported a failed outburst occurring in the BH transient Swift~J1745--26 with the source brightness up to $0.4~L_{\rm EDD}$, implying that a low luminosity is not the sole criterion for such a phenomenon. 
Adopting a distance to the source of 3~kpc \citep{Atri2019}, a BH mass of 9~$M_{\rm \odot}$ \citep{Atri2019} and the maximum unabsorbed flux of $4 \times 10^{-8}~\rm erg~s^{-1}~cm^{-2}$ from the fit to the {\it NICER} spectrum in the 1--10~keV band, the peak luminosity of {\sub} in our work is around $0.04~L_{\rm EDD}$, inconsistent with the case reported by \cite{DelSanto2016}.

To further investigate this failed outburst in {\sub}, we conducted spectral analysis using {\it NICER} data between MJD~58198. 
The changes in the disk temperature and the photon index between epoch~1 and 2 indicate that the source first softened and then hardened. It is worth noting that the inner disk radius decreased as the disk temperature increased in epoch~1 whereas the inner disk radius remained constant while the disk temperature decreased in epoch~2; the inner disk radius in epoch~2 is overall smaller than that in epoch~1. These changes suggest that the accretion disk moves inwards even when the spectrum is dominated by the hard component, and it stops at a position much larger than the innermost stable circular orbit (adopting a mass of the black hole from \citealt{Atri2019}, $M=9~M_{\rm  \odot}$, assuming $R_{\rm ISCO}=6~R_{\rm g}=6~GM/c^2$, $R_{\rm ISCO}= 79.7$~km). 
In addition, the changes of the disk temperature and the inner disk radius in epoch~2 are inconsistent with the prediction of the standard accretion disk model, implying that some other physical process(es) leading to these changes is(are) yet to be known.

Both \cite{Kara2019} and \cite{Buisson2019} suggest that the inner part of the accretion disk in {\sub} did not evolve in the hard state of this outburst. However, if we only look into the period of epoch~1, the inner radius of the accretion disk measured by \cite{Buisson2019} decreases with time as well, which is consistent with our result.
However, the best-fitting disk inner radius, also the photon index reported here is larger than that in, for instance, \cite{Buisson2019} and You. et al (in prep) and thus the absolute values of the spectral parameters reported in this work should be taken carefully.

\subsection{High-frequency time lags associated with photon index}
We studied the frequency-dependent time lags between the soft, 1--5~keV, and the hard energy bands, up to 150~keV. 
The low-frequency lags show both/either soft and/or hard lag in one observation, and appear to be independent of energy even though the lag between the 1--5~keV and the 80--150~keV is in average larger than the ones between the other energy bands. The high-frequency lags, however, show clear evolution along the outburst: they first increase along the outburst in epoch~1 and then decrease in epoch~2, and they are energy-dependent, with the magnitudes of the lags increasing energy (see Fig.~\ref{fig:lag}).

Fig.~\ref{fig:spec} shows that the spectra in both epochs are dominated by the hard emission, and that the {\sc diskbb} flux only contributes 4--6\% to the total flux in the 1--10~keV band. 
As shown in Fig.~\ref{fig:lag_gamma}, the evolution of the high-frequency lags is highly correlated to that of the photon index of the hard spectral component, hinting that the lags are associated with the Comptonized component.
The observed high-frequency time lags are thus largely produced by the hard photons, i.e. the high-frequency time lags mainly take place in the Comptonizing region.

\cite{Miyamoto1988} observed a period-dependent time-lag behavior in Cyg~X--1, and they proposed that such phenomenon is due to a perturbing wave travelling from a low energy- to a high-energy-X-ray-emitting region in the accretion disk and the time delay corresponds to the wave travel-time. \cite{Cabanac2010} further proposed that the broadband noise and low-frequency QPO in the power spectrum could be produced by a magneto-acoustic wave propagating within the corona. Following equation (1) in \cite{Cabanac2010}, the speed of the sound wave is $c_{\rm s} \simeq 3.1\times 10^{8}~( \frac{T_{0}}{\rm 100 keV} )^{1/2}~\rm cm~s^{-1}$, proportional to the square root of the electron temperature of the corona. Although the energy band of the {\it NICER} data does not allow us to constrain the coronal temperature in {\sub}, \cite{Buisson2019} shows the electron temperature in the corona in {\sub} is proportional to the photon index during the period used in this work. Moreover, \cite{Kara2019} demonstrated that the corona in {\sub} shrinks vertically in epoch~1. The combination of these two facts, increasing sound speed and decreasing path, would lead to a decreasing time lag in this period. Our observations show the opposite behavior. This rules out the possibility that the high-frequency time lag in {\sub} is only caused by a sound wave traveling within the corona.

Another potential model to explain this phenomenon is the inverse Comptonization of soft photons by energetic electrons in a jet \citep{Reig2018}. In their model, when the optical depth/electron density is high, the soft photons are inversely Comptonized by the energetic electrons in the base of the jet. Because of the short mean free path, hard photons are scattered on a short length-scale and hence the corresponding time lag between soft and hard photons is small. As the optical depth decreases, the mean free path increases, leading to a longer time lag and a softer spectrum. 
In sum, the average time lag of the hard photons with respect to the soft ones increases as the X-ray emission becomes softer.
Steady compact radio jets in the hard state of {\sub} have been confirmed by \cite{Atri2019}, supporting the model proposed by \cite{Reig2018} as a possible explanation of the high-frequency time lags observed in {\sub}. However, the observed time lags are actually smaller than the ones reported by \cite{Reig2018} in similar energy bands. Especially, it may be difficult for this model to produce the factor of $\sim$10 larger lags in the 1--10~Hz frequency range when the hard photon energy increases from several keV to about 100~keV.

A reverberation lag occurs when some of the coronal photons irradiating the accretion disk get reflected and lag behind the primary photons (see a review of the reverberation lag in \citealt{Uttley2014}). 
The reverberation lag is thus soft and can constrain the distance between the hard emission region and the accretion disk.
\cite{Kara2019} observed a few millisecond reverberation lag between the 0.5--1~keV and 1--10~keV bands decreasing with time while the accretion disk remains unchanged in {\sub}, and they suggested that the corona contracts in this period.
Compared to their result, the 1--10~Hz time lags observed in this work are hard and much larger. Although, as a result of the inverse Comptonization in a jet, such lags indicate a large emission region, up to 1000~$R_{\rm g}$, this is different from the distance of about $\sim 10~R_{\rm g}$ between the accretion disk and the compact corona reported by \cite{Kara2019}.
However, by jointly fitting the observed motion of the jets in radio and X-rays, \cite{Espinasse2020} obtained an ejection date of the mater in the jets at around MJD~58305, in which the jet size is very large, i.e. $1.5 \times 10^4$~AU by $7.7 \times 10^3$~AU. This result is consistent with a jet of a scale of 1000~$R_{\rm g}$ in earlier observations.
Overall, the results in \cite{Kara2019} and our work suggest that there are two hard emission regions during the studied period of the outburst in {\sub}: one is a compact corona and another one is a jet with a large scale.

For the low-frequency time lags, the timescale, dozens of seconds, is more comparable to the time lag predicted by the model of \cite{Kotov2001} and \cite{Arevalo2006}. They explain the lags as the result of viscous propagation of mass accretion fluctuations within the inner regions of the disk.

\section{Conclusion}
We study the evolution of the temporal and spectral properties of {\sub} in its luminous hard state with {\hx} and {\it NICER} and find that (1) a hard-to-hard transition occurred at around MJD~58257, making {\sub} a new case of a transient displaying a failed outburst; (2) the broadband fractional rms amplitude and time lags are all frequency-, flux- and energy-dependent; (3) the 1--10~Hz time lags are correlated with the photon index of the hard spectral component, qualitatively supporting the model of the lags as caused by Comptonization in a jet.

\emph{Acknowledgements}
We thank the referee for the useful feedback for improving the manuscript.
This work made use of the data from the {\hx} mission, a project funded by China National Space Administration (CNSA) and the Chinese Academy of Sciences (CAS). The {\hx} team gratefully acknowledges the support from the National Program on Key Research and Development Project (Grant No. 2016YFA0400800) from the Minister of Science and Technology of China (MOST) and the Strategic Priority Research Program of the Chinese Academy of Sciences (Grant No. XDB23040400). The authors thank supports from the National Natural Science Foundation of China under Grants No. 11673023, 11733009, 11603037, 11973052, U1838201, U1838115, U1938103 and U1838202.

\bibliographystyle{aasjournal}
\bibliography{ref_2019}

\appendix
\renewcommand{\thefigure}{\hbAppendixPrefix\arabic{figure}}

\numberwithin{figure}{section}

\section{Power spectra and the corresponding time lags in {\sub}} 
\begin{figure}
\mbox{\includegraphics[width=7cm]{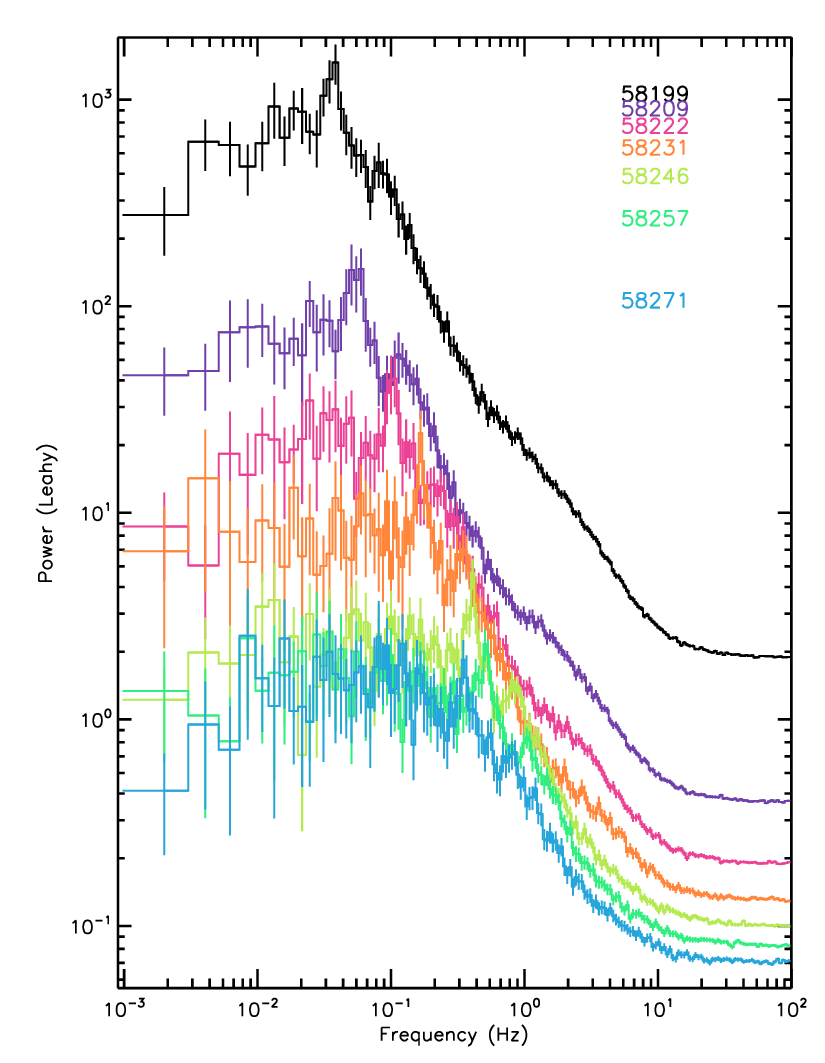}}
\hspace{-.1cm}
\mbox{\includegraphics[width=7cm]{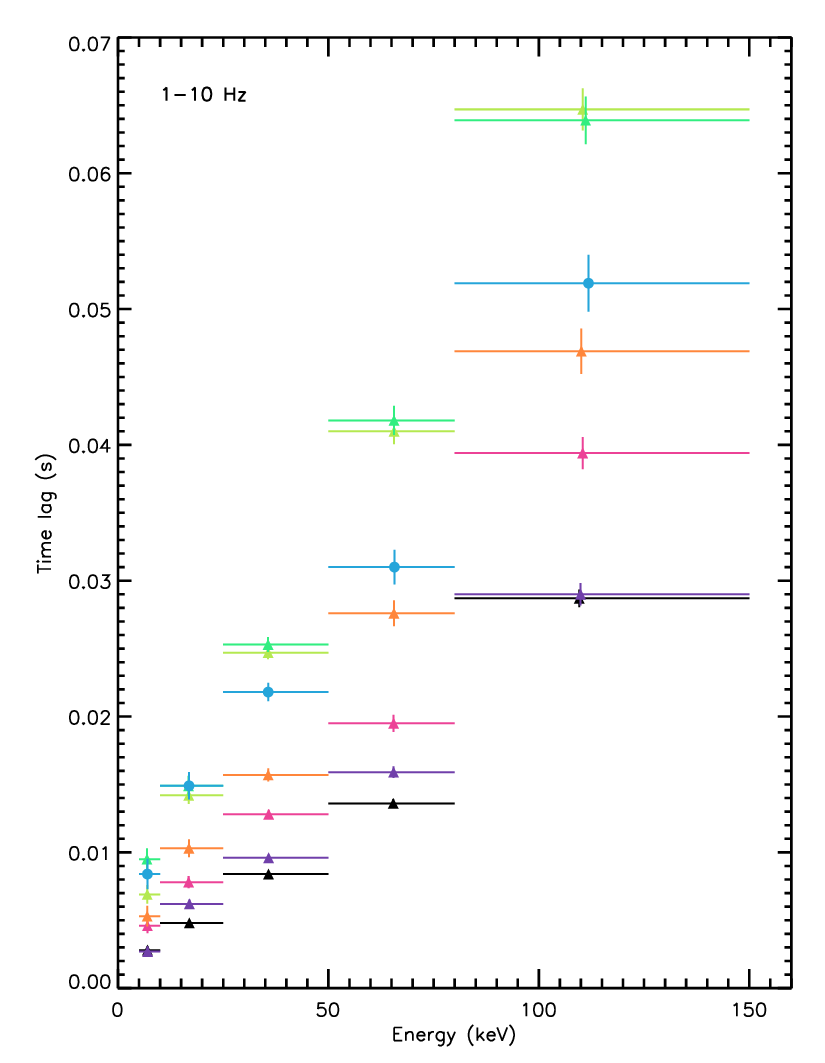}}

\caption{Left: power spectra in {\sub}. Except for the PDS at 58199, the following ones have been re-scaled by multiplying a factor of 0.2, 0.1, 0.067, 0.05, 0.033 and 0.029, respectively, for displaying purpose. The legends on the top right of the panel indicate the observation time of each PDS in MJD.
Right: the 1--10~Hz time lags of the shown energy bands with respect to the 1--5~keV band, corresponding to the power spectra in the left panel. The triangles and circle correspond to epoch~1 and 2, respectively.} 
\label{fig:lag_obs}
\end{figure}

\end{document}